\documentclass[%
twocolumn,
superscriptaddress,
showpacs,
nofootinbib,
 amsmath,amssymb,
 aps,
]{revtex4-1}

\usepackage{graphicx}
\usepackage{dcolumn}
\usepackage{bm}
\usepackage{multirow}
\usepackage{hyperref}

\usepackage{braket}
\usepackage{mathtools}
\usepackage[usenames,dvipsnames,svgnames,table]{xcolor}
\usepackage [utf8] {inputenc}
\usepackage{color}
\usepackage{subfig}
\usepackage{lmodern}
\usepackage{footnote}
\usepackage{slashed}
\usepackage{verbatim}

\newcommand{\cD}[0]{\mathcal D}
\newcommand{\cK}[0]{\mathcal K}
\newcommand{\cL}[0]{\mathcal L}
\newcommand{\cM}[0]{\mathcal M}
\newcommand{\cO}[0]{\mathcal O}

\newcommand{\cS}[0]{\mathcal S}

\newcommand{\wt}[0]{\widetilde}

\newcommand{\df}[0]{\mathrm{df}}
\newcommand{\iso}[0]{{\rm iso}}

\newcommand{\Kiso}[0]{{\cK_{\df,3}^{\iso}}}
\newcommand{\Kisozero}[0]{{\cK_{\df,3}^{\iso,0}}}
\newcommand{\Kisoone}[0]{{\cK_{\df,3}^{\iso,1}}}
\newcommand{\Kdf}[0]{{\cK_{\df,3}}}

\newcommand{\PV}[0]{{\mathrm{PV}}}
\newcommand{\Mdf}[0]{\mathcal{M}_{\df,3}}
\newcommand{\K}[0]{\mathcal K}

\newcommand{\HSTH}[0]{Hansen:2016fzj}
\newcommand{\BHSQC}[0]{Briceno:2017tce}
\newcommand{\BHSnum}[0]{Briceno:2018mlh}
\newcommand{\BHSK}[0]{Briceno:2018aml}
\newcommand{\HSQCa}[0]{Hansen:2014eka}
\newcommand{\HSQCb}[0]{Hansen:2015zga}
\newcommand{\dwave}[0]{Blanton:2019igq}

\newcommand{\Akakia}[0]{Hammer:2017uqm}
\newcommand{\Akakib}[0]{Hammer:2017kms}

\newcommand{\MD}[0]{Mai:2017bge}
\newcommand{\MDpi}[0]{Mai:2018djl}
\newcommand{\Akakinum}[0]{Doring:2018xxx}

\newcommand{\HSrev}[0]{Hansen:2019nir}

\newcommand{\HH}[0]{Horz:2019rrn}
\newcommand{\largera}[0]{Romero-Lopez:2019qrt}

\newcommand{\chpt}{$\chi$PT}
\newcommand{\p}{\partial }

\newcommand{\tr}{\text{tr} \!}

\begin{document}


\title{$I=3$ three-pion scattering amplitude from lattice QCD}

\author{Tyler D. Blanton}
\email{blanton1@uw.edu}
\affiliation{Physics Department, University of Washington, Seattle, WA 98195-1560, USA}
\author{Fernando Romero-L\'opez}
\email{fernando.romero@uv.es}
\affiliation{Instituto de F\'isica Corpuscular, Universitat de Val\`encia and CSIC,
46980 Paterna, Spain}
\author{Stephen R. Sharpe}
\email{srsharpe@uw.edu}
\affiliation{Physics Department, University of Washington, Seattle, WA 98195-1560, USA}


\date{\today}

\begin{abstract}
We analyze the spectrum of two- and three-pion states of maximal isospin obtained recently
for isosymmetric QCD with pion mass $M\approx 200\;$MeV in Ref.~\cite{\HH}.
Using the relativistic three-particle quantization condition, we find $\sim 2 \sigma$ evidence
for a nonzero value for the contact part of the $3\pi^+$ ($I=3$) scattering amplitude.
We also compare our results to leading-order chiral perturbation theory. We find good agreement at threshold, and some tension in the energy dependent part of the $3\pi^+$scattering amplitude.
We also find that the $2\pi^+$ ($I=2$) spectrum is fit well by an $s$-wave phase shift
that incorporates the expected Adler zero. 
\end{abstract}

\maketitle


\section{\label{sec:intro}Introduction}

Lattice QCD (LQCD) provides a powerful (if indirect) tool for \textit{ab initio} calculations of
strong-interaction scattering amplitudes.
The formalism for determining two-particle amplitudes is
 well understood~\cite{Luscher:1986n2,Luscher:1991n1,Kari:1995, Kim:2005gf, He:2005ey, Bernard:2010, Hansen:2012tf, Briceno:2012yi, Briceno:2014oea, Romero-Lopez:2018zyy,Luu:2011ep,Gockeler:2012yj},
 and there has been enormous progress in its implementation in recent years~\cite{Feng:2009ij,Lage:2009zv,Wilson:2015dqa, Briceno:2016mjc, Brett:2018jqw, Andersen:2017una, Guo:2018zss, Andersen:2018mau, Dudek:2014qha, Dudek:2016cru, Woss:2018irj, Woss:2019hse,Helmes:2018nug,Liu:2016cba,Helmes:2017smr,Helmes:2015gla,Werner:2019hxc,Culver:2019qtx,Mai:2019pqr,Doring:2012eu} 
 (see Ref.~\cite{Briceno:2017max} for a review).
 The present frontier is the determination of three-particle scattering amplitudes
 and related decay amplitudes. 
 LQCD calculations promise
 access to three-particle scattering processes that are difficult or impossible
  to access experimentally.
 Examples of important applications
 are understanding properties of resonances with significant three-particle branching
 ratios (including the Roper resonance~\cite{Roper}, 
 and many of the $X$, $Y$ and $Z$ resonances~\cite{Lebed:2016hpi}),
 determining the three-nucleon interaction (important for large nuclei and neutron star properties),
predicting  weak decays to three particles (e.g. $K\to 3\pi$),
and calculating the $3\pi$ contribution to the hadronic-vacuum polarization that enters
into the prediction of muonic $g-2$~\cite{Hoferichter:2019gzf}.

Three-particle amplitudes are determined using LQCD by calculating the energies of two- and 
three-particle states in a finite volume \cite{Polejaeva:2012ut,Kreuzer:2012sr}. The challenges to carrying this out are twofold. On the one hand,
the calculation of spectral levels becomes more challenging as the number of particles increases.
On the other, one must develop a theoretical formalism relating the spectrum to scattering amplitudes.
Significant progress has recently been achieved in both directions, with energies well above the 
three-particle threshold being successfully measured, 
and a formalism for three identical (pseudo)scalar particles available. 
The formalism has been developed and implemented following three approaches: 
generic relativistic effective field theory 
(RFT)~\cite{\HSQCa,\HSQCb,\BHSQC,\BHSnum,\BHSK,\dwave,\largera},
nonrelativistic effective field theory (NREFT)~\cite{\Akakia,\Akakib,\Akakinum,Pang:2019dfe}, 
and (relativistic) finite volume unitarity (FVU)~\cite{\MD,\MDpi} (see also Refs. \cite{Klos:2018sen,Guo:2017ism} and Ref.~\cite{\HSrev} for a review).
To date, only the RFT formalism has been explicitly worked out including
higher partial waves.
The application to LQCD results has so far been restricted to the energy of
the three-particle ground state, either using the threshold 
expansion~\cite{Beane:2007qr,Detmold:2008fn,Romero-Lopez:2018rcb}, or, more recently, the FVU approach for $3 \pi^+$~\cite{Mai:2018djl}. 

Recently, precise results were presented for the spectrum of
$2\pi^+$ and $3 \pi^+$ states in $O(a)$-improved 
isosymmetric QCD with pions having close to
physical mass, $M \approx 200\;$MeV~\cite{\HH}.
These were obtained in a cubic box of length $L$ with $M L \approx 4.2$,
for several values of the total momentum $\vec P=(2\pi/L)\vec d$ with $\vec d \in \mathbb{Z}^3$, and for several
irreducible representations (irreps) of the corresponding symmetry groups.
Isospin symmetry ensures that G parity is exactly conserved and thus
that the $2\pi^+$ and $3\pi^+$ sectors are decoupled.
In total, sixteen $2\pi^+$ levels and eleven $3\pi^+$ levels were obtained below the
respective inelastic thresholds at $E_2^*=4 M$ and $E^*=5 M$,
Here $E_2^*$ and $E^*=\sqrt{E^2-\vec P^2}$ are the corresponding center-of-mass energies,
with $E$ the total three-particle energy.

The purpose of this Letter is to perform a global analysis of
the spectra of Ref.~\cite{\HH} using the RFT formalism
and determine the underlying $3\pi^+$ interaction.
This breaks new ground for an analysis of the three-particle spectrum in several ways:
we use multiple excited states, in both trivial and nontrivial irreps, including results from moving frames.
This analysis therefore serves as a testing ground for the utility of the three-particle formalism in an almost
physical example.
An additional appealing feature is that the size of the $3\pi^+$ interaction can be calculated
using chiral perturbation theory (\chpt). We present the leading order (LO) prediction here.

After this paper was made public, an independent study of the results of Ref. \cite{\HH}, using the FVU approach, appeared \cite{Mai:2019fba}.

\section{Formalism and Implementation\label{sec:formalism}}
All approaches to determining three-particle scattering amplitudes using LQCD proceed
in two steps, which we outline here.
In the first step, one uses a quantization condition (QC), which predicts
the finite-volume spectrum in terms of an intermediate infinite-volume three-particle scattering quantity.
In the RFT approach, the QC for identical, spinless particles with a G-parity-like
$Z_2$ symmetry takes the form
(up to corrections of $\cO(1\%)$ that are exponentially-suppressed in $ML$)~\cite{\HSQCa} 
\begin{equation}
\det\left[ F_3(E,\vec P, L)^{-1}+ \Kdf(E^*) \right] = 0\,.
\label{eq:QC}
\end{equation}
Here $F_3$ and $\Kdf$ are matrices in a space describing three on-shell particles in finite volume. They have indices of angular momentum of the interacting pair, $\ell,m$, and finite-volume momentum of the spectator particle, $k$. $F_3$ depends on the two-particle scattering amplitude and on known geometric functions,
while $\Kdf$ is the three-particle scattering quantity referred to above.
It is quasilocal, real, and free of singularities related to three-particle thresholds, thus playing a similar
role to the two-particle K matrix $\cK_2$ in two-particle scattering. 
It is, however, unphysical, as it  depends on an ultraviolet (UV) cutoff.
Given prior knowledge of $\cK_2$, and a parametrization of $\Kdf$, the energies of finite-volume states
are determined by the vanishing of the determinant in Eq.~(\ref{eq:QC}).
The parameters in $\Kdf$ are then adjusted to fit to the numerically-determined spectrum. Examples on how to numerically solve Eq. \eqref{eq:QC} have been presented in Refs. \cite{\BHSnum, \dwave,\largera}.

The second step requires solving infinite-volume integral equations in order to  relate $\Kdf$ to the
three-particle scattering amplitude $\cM_3$. In fact, as explained below, it is a divergence-free
version of the latter, denoted $\Mdf$, that is most useful. The equations relating $\Kdf$ to $\Mdf$
were derived in Ref.~\cite{\HSQCb}, and solved in Ref. \cite{\BHSnum}.

The parametrizations we use for $\cK_2$ and $\Kdf$ are based on an expansion about two- and 
three-particle thresholds. For $\cK_2$ this leads to the standard effective range expansion (ERE), 
recalled below. At linear order in this expansion only $s$-wave interactions are nonvanishing,
with $d$-wave interactions first entering at quadratic order
($p$-wave interactions are forbidden by Bose symmetry).
For $\Kdf$, the expansion is in powers of
$\Delta = (E^{*2}- 9 M^2)/ (9 M^2)$, and was developed in Refs.~\cite{\BHSnum,\dwave}
based on the Lorentz and particle-interchange invariance of $\Kdf$.
Through linear order in $\Delta$, $\Kdf$ is given by 
\begin{align}
\Kdf = \Kiso =\Kisozero + \Kisoone  \Delta \,,
\label{eq:threxp}
\end{align} 
where $\Kisozero$ and $\Kisoone$ are constants.
There is no dependence on the momenta of the three particles at this order;
this corresponds to a contact interaction, and leads to the designation ``isotropic''.
Momentum dependence first enters at $\cO(\Delta^2)$. 

In our main analysis we keep only the $s$-wave two-particle interaction and the isotropic terms
in Eq.~(\ref{eq:threxp}).
With these approximations, the QC of Eq.~(\ref{eq:QC}) reduces to a finite matrix equation that can
be solved by straightforward numerical methods. Previous implementations have considered only
the three-particle rest frame, $\vec P=0$~\cite{\BHSnum,\dwave,\largera}
(see also Ref.~\cite{\Akakinum,\MDpi}). Here we have extended the
implementation to moving frames, so that we can use all the results obtained by Ref.~\cite{\HH}.
The details of the implementation, including projections onto irreps
of the appropriate little groups, are described in Appendix~\ref{app:moving}.


\section{\chpt\ prediction for $\Kdf$ and $\Mdf$\label{sec:chpt}}


$\Mdf$ and $\Kdf$ have not previously been calculated in \chpt, so here we present
the leading order (LO) result. The LO Lagrangian in the isosymmetric two-flavor theory is \cite{Weinberg:1978kz,Gasser:1983yg} 
\begin{align}
\begin{split}
&\mathcal{L_\chi} = \frac{F^2}{4} \tr \left( \p_\mu U  \p^\mu U^\dagger  \right) 
+ \frac{M^2 F^2}{4 } \tr \left( U  + U^\dagger \right)\,,\\ 
&\text{ with  } U = e^{i\phi/F} 
\text{ ~and~ }
\phi = \begin{pmatrix}
\pi^0 & \sqrt{2} \pi^+ \\
\sqrt{2} \pi^- & -\pi^0 
\end{pmatrix}\,.
\end{split}
\end{align} 
Here $F$ is the decay constant in the chiral limit, normalized such that $F_\pi=92.4\;$MeV. We note that at this order, $F=F_\pi$.
Expanding in powers of the pion fields, 
$\mathcal{L} = \mathcal{L}_{2 \pi}  + \mathcal{L}_{4 \pi} + \mathcal{L}_{6 \pi} + \cdots,$
we need only the $4\pi$ and $6\pi$ vertices.

From $ \mathcal{L}_{4 \pi}$ we obtain the standard LO result for
the $2\pi^+$ scattering amplitude \cite{Weinberg:1966kf},
\begin{equation}
\mathcal{M}_2 =  \frac{2M^2 - E_2^{*2}}{F^2} \,,
\label{eq:M2}
\end{equation}
which displays the well-known Adler zero 
below threshold at $E_2^{*2}= 2 M^2$~\cite{Adler:1964um}. 
Given the ERE parametrization of the $s$-wave phase shift,
\begin{equation}
q \cot \delta_0(q) = - \frac{1}{a_0} + \frac{r q^2}{2}	 + P r^3 q^4 + \cdots \label{eq:ERE}\,,
\end{equation}
where $q^2=E_2^{*2}/4-M^2$,
one can infer from Eq.~(\ref{eq:M2}) the LO results for the scattering length and effective range:
\begin{equation}
 M a_0  = \frac{M^2}{16 \pi F^2}\ \text{ and } \  M^2 r a_0 = 3\,.
 \label{eq:chptLO}
\end{equation}

\begin{figure}[h!]
\begin{center}
\includegraphics[width=.40\textwidth]{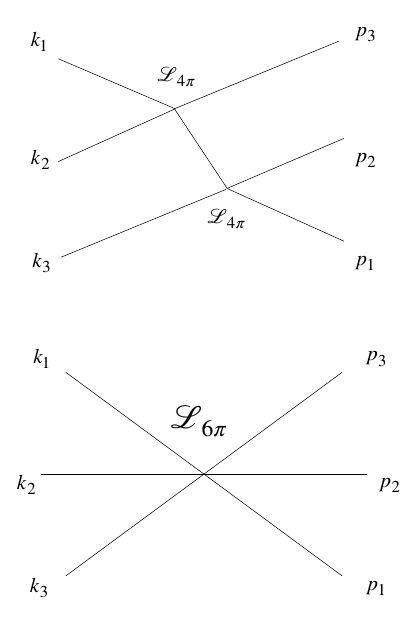}
\caption{LO contributions to the three-particle scattering amplitude $\mathcal{M}_3$.
Momentum assignments must be symmetrized.}
\label{fig:diags}
\end{center}
\end{figure}

The $3\pi^+$ amplitude $\cM_3$ is given at LO by the diagrams of Fig.~\ref{fig:diags}. 
As is well known, $\cM_3$ diverges for certain external momenta, 
as the propagator in Fig.~\ref{fig:diags}(a) can go on shell.
This motivated the introduction of a divergence-free amplitude in Ref.~\cite{\HSQCa}:
\begin{align}
\ \ \ \ \ \ \Mdf &\equiv \mathcal{M}_3 - \mathcal{D}\,,
\label{eq:Mdf3}
\\
\mathcal{D} = \cS\Big\{ - \mathcal{M}_2(s_{12}) &\frac{1}{b^2-M^2} \mathcal{M}_2(s'_{12}) \Big\}
+ O\left( \mathcal{M}_2^3 \right)\,,
\label{eq:D}
\end{align}
where $s_{12}=(p_1+p_2)^2$, $s'_{12}=(k_1+k_2)^2$,  $b=p_1+p_2-k_3$,
and $\cS$ indicates symmetrization over momentum assignments.
$\mathcal{D}$ is defined to have the same divergences as $\mathcal{M}_3$,
so that their difference is finite.
At LO in \chpt, only the LO term in $\cD$ contributes and we find
\begin{align}
\begin{split}
M^2 \Mdf &= \frac{M^4}{F^4}(18 + 27 \Delta) \\ &= (16\pi M a_0)^2 (18+27\Delta)\,,
\label{eq:Mdfres}
\end{split}
\end{align}
a result that is real and isotropic. 
As a side result, we have also calculated the related threshold amplitude that enters into the
$1/L$ expansion of the three-particle energy~\cite{\HSTH}, finding
$\cM_{3,\rm th}= 27 M^2/F^4$.

The last step is to relate $\Mdf$ to $\Kdf$. As discussed in Appendix~\ref{app:nonleading},
we find these quantities to be equal at LO
\begin{equation}
\Kdf = \Mdf \left[1 + \mathcal{O}(M^2/F^2) \right]\,,
\label{eq:KtoM0}
\end{equation}
so that $\Kdf$ is also given by Eq.~(\ref{eq:Mdfres}). 
This implies that $\Kdf$ is scheme-independent at LO in \chpt. 
In Appendix~\ref{app:nonleading} we also quantify the expected size of the corrections, 
finding them to range between $10-50\%$, with the larger error applying to the
term linear in $\Delta$.

\section{Fitting the two-particle spectrum\label{sec:twopart}}

Determining the two-particle phase shift is an essential step,
as it enters into the three-particle QC.
In particular, we need a parametrization valid below threshold, 
as the two-particle momentum in the three-particle QC takes values in the range $q^2/M^2 \in [-1,3]$.
We extract information on the $s$-wave phase shift using 
a form of the two-particle QC that holds in all frames for those irreps that couple to $J=0$.
Details are given in Appendix~\ref{app:fits}.
We use the bootstrap samples provided in
Ref.~\cite{\HH} to determine statistical errors, so that correlations are accounted for properly.

We use a parametrization of the phase shift 
(adapted from that of Ref.~\cite{Yndurain:2002ud}; see also Ref.~\cite{Pelaez:2019eqa}) 
that includes the Adler zero  predicted by \chpt, as well as the kinematical factor $E_2^*$:
\begin{multline}
\frac{q}{M} \cot \delta_0(q) =\\
\frac{E_2^* M}{E_2^{*2}\!-\!2z_2^2} 
\left( B_0 \!+\! B_1 \frac{q^2}{M^2} \!+\! B_2 \frac{q^4}{M^4} \!+\! \cdots \right)\,. \label{eq:Adler}
\end{multline}
We either set $z_2^2=M^2$, the LO value, or leave it as a free parameter.
$B_0$ and $B_1$ are related in a simple way to $a_0$ and $r$
[see Eqs.~(\ref{eq:a0Adler}) and (\ref{eq:ra0Adler}) in Appendix~\ref{app:fits}].
{
Previous lattice studies have used the ERE, Eq.~(\ref{eq:ERE})
(see, e.g. Refs.~\cite{Beane:2011sc,Dudek:2012gj,Bulava:2016mks}), but this has
the disadvantage, due to the Adler zero, of
having a radius of convergence of $|q^2|=|M^2-z_2^2/2|\approx M^2/2$.
In particular, the ERE gives results for $-1 < q^2/M^2 < 0$ that are substantially different
from the Adler-zero form.
This is related to the fact that in (\ref{eq:Adler}), $B_1$ and $B_2$ are both of
next-to-leading order (NLO) in \chpt, 
 in contrast to the ERE form where $r$ and $P$ are both nonzero at LO,
as can be seen from the explicit \chpt\ expressions given in Ref.~\cite{Beane:2011sc}.
The formal radius of convergence of our expression (\ref{eq:Adler}) is $|q^2|= M^2$,
due to the left-hand cut, but following common practice we ignore this and use it
up to $q^2/M^2=3$. 
In Appendix~\ref{app:fits} we show that fitting with the restriction
$|q^2|/M^2 < 1$ has only a small impact on the resulting parameters
We also have checked that
fits using the ERE form provide a worse description of the data.}

\begin{figure}[h!]
\begin{center}
\includegraphics[width=.48\textwidth]{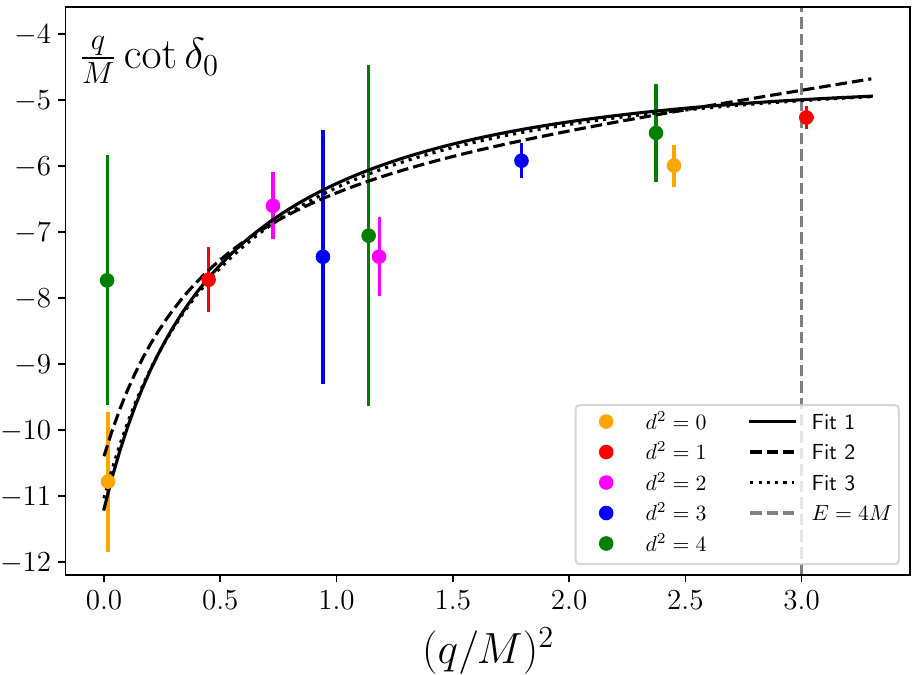}
\caption{Values of $q\cot\delta_0$ obtained from the two-particle spectrum of Ref.~\cite{\HH} using
the two-particle QC, together with various fits.
}
\label{fig:delta0}
\end{center}
\end{figure}

\begin{table*}[t]
\centering
\begin{tabular}{cccccccc}
Fit                         & $B_0$ & $ B_1$ & $B_2$ & $z^2_2/M^2$ & $\chi^2/$dof & $M a_0$ & $M^2 r a_0$ \\ \hline\hline
1 &   -11.2(7) &  -2.1(3)         &   ---    &    1 (fixed)     &    12.13/(11-2)  &\  0.089(6) \  &  2.63(8)  \\ \cline{1-8} 
 2                           &  -10.4(9) & -3.7(1.0)&   0.5(3)     &  1 (fixed)      & 9.75/(11-3)    & \ 0.096(8) \ & 2.3(3)    \\ \cline{1-8} 
  3                         &   -11.7(1.8) &  -2.0(4)   &--- &0.94(22)      &     12.06/(11-3)  &\  0.091(9) \ &  2.4(9) 
\end{tabular}
\caption{Fits of the two-particle spectrum to the Adler-zero form of $q\cot\delta_0$, Eq.~(\ref{eq:Adler}).
}
\label{tab:resAdler}
\end{table*}

{
The results of several fits are listed in Table~\ref{tab:resAdler} and
shown in Fig.~\ref{fig:delta0}. All fits give reasonable values of
$\chi^2/\text{d.o.f.}$, and yield values for $M^2 r a_0$ close to the
predicted LO value of $3$.
Using the value of $F$ obtained from the same lattice
configurations in Ref.~\cite{Bruno:2016plf, Bruno:2014jqa}, 
the LO chiral prediction from Eq.~(\ref{eq:chptLO})  is $M a_0=0.0938(12)$, 
and this is also in good agreement with the results of the fits.
Overall, we conclude that the spectrum from Ref.~\cite{\HH} confirms the expectations from \chpt.
We choose the minimal fit 1 as our standard choice since $B_2$ is poorly determined (fit 2) and 
the Adler-zero position is consistent with the LO result if allowed to float (fit 3).
}
We have performed a similar fit to the five energy levels from Ref.~\cite{\HH} 
which are sensitive only to the $d$-wave amplitude. 
Details are in Appendix~\ref{app:fits}.
Despite very small shifts
from the free energies, we find a 3$\sigma$ signal for the $d$-wave scattering length,
$(Ma_2)^5=0.0006(2)$, 
where $a_2$ is defined in Eq.~\eqref{eq:a2} of Appendix~\ref{app:fits}.
The smallness of this result is qualitatively consistent with the fact that this is a NLO effect in \chpt,
and justifies our neglect of $d$-waves in the three-particle analysis.

\section{Fitting the three-particle spectrum\label{sec:threepart}}

We now use the three-particle spectrum to determine $\Kiso$. Eight levels are sensitive to $\Kiso$,
while three are in irreps only sensitive to two-particle interactions. 
Since all levels are correlated, a global fit to 
two- and three-particle spectra is needed to properly estimate errors.
Further details on the fits described in this section can be found in Appendix~\ref{app:fits}.

\begin{figure}[h!]
\begin{center}
\includegraphics[width=.48\textwidth]{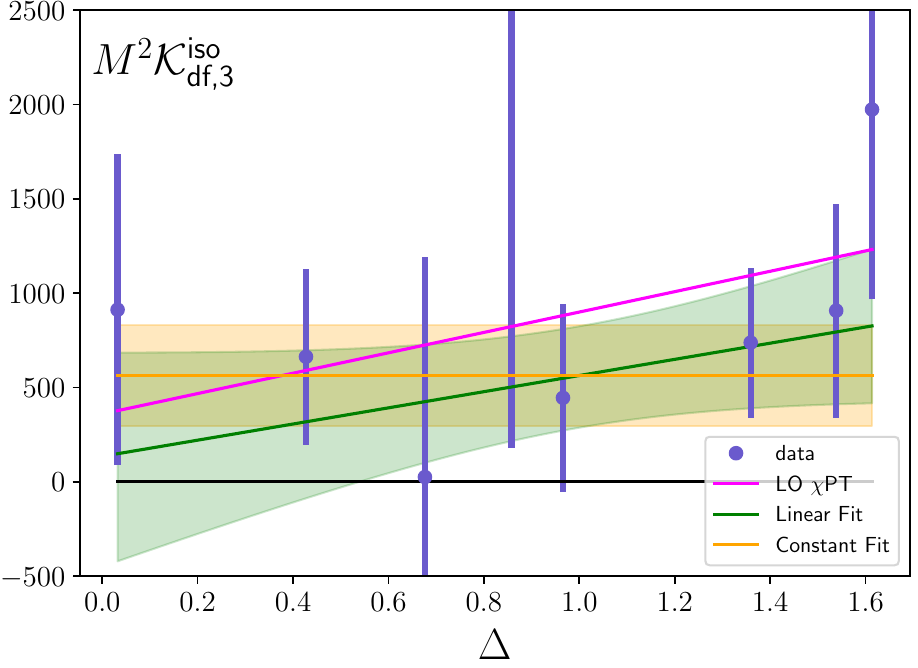}
\caption{Results for $M^2 \Kiso$ from individual three-particle levels, using method 1,
together with constant and linear fits, and the LO prediction of \chpt.
}
\label{fig:Kiso}
\end{center}
\end{figure}

 Before presenting the global fits, however, 
 we use an approach (``method 1'') that
 allows a separate determination of $\Kiso$ for each of the eight levels sensitive to this parameter.
 Within each bootstrap sample, we fit the two-particle levels to the fit 1 Adler-zero form described
 above, and then adjust $\Kiso$ so that the three-particle QC reproduces the energy of the
 level under consideration.
 The results are shown in Fig.~\ref{fig:Kiso}. The values of $\Kiso$ are all positive,
and a constant fit yields $M^2\Kiso=560(270)$ with $\chi^2/{\rm d.o.f.}= 8.5/7$.
The LO \chpt\ result (given by $M^2\Kiso=360+540\Delta$,  taking $M a_0$ from fit 1) 
is reasonably consistent with the linear fit, as shown. 
This indicates that a significant result for $\Kiso$ of the expected size may be obtainable. 

This fit does not include three-particle energy levels in irreps sensitive only to $\delta_0$.
These, however, can be used as a consistency check. 
As shown in Appendix~\ref{app:fits},
we find good agreement between the data and the energies predicted by the QC.

\begin{table*}[t]
\centering
\begin{tabular}{ccccccccc}
Fit                         & $B_0$ & $ B_1$ & $z^2_2/M^2$ & $M^2\Kisozero$ & $M^2\Kisoone$ & $\chi^2/$dof & $M a_0$ & $M^2 r a_0$ \\ \hline\hline
4 &  -11.1(7) & -2.3(3)&  1  (fixed)  &  270(160) & ---    & 27.06/(22-3)    & 0.090(6) & 2.59(8)    
\\ \hline
5  &   -11.1(7) &  -2.4(3)   & 1  (fixed) & 550(330) & -280(290)   & 26.04/(22-4)  & 0.090(5) & 2.57(8)
\end{tabular}
\caption{Global fits to the two- and three-particle spectrum using the two- and three-particle QCs.
}
\label{tab:global}
\end{table*}

To establish the true significance of the results for $\Kiso$ we perform global fits to the eleven two-particle and eleven three-particle levels that depend on $\delta_0$ and/or $\Kiso$.
We do so both for constant and linear $\Kiso$. The results are collected in Table~\ref{tab:global}.
Fit 4 finds a value for $\Kiso$ that has around 1.8$\sigma$ {statistical} significance,
and also gives values for $B_0$ and $B_1$ that are consistent with those from fits 1-3 above
and with the LO \chpt\ predictions.  The p-value of the fit is $p=0.103$.

In fit 5, we try a linear ansatz for $\Kiso$, and find that the current dataset of Ref.~\cite{\HH} 
is insufficient for a separate extraction of both constant and linear terms. 
We  note, however, that, even in this fit, the scenario
$\Kiso=0$ is excluded at $\sim 2\sigma$.  We also provide a comparison between the data and the predicted spectrum from this fit in Fig. \ref{fig:spectra} of Appendix \ref{app:fits}.

\begin{figure}[h!]
\begin{center}
\includegraphics[width=.48\textwidth]{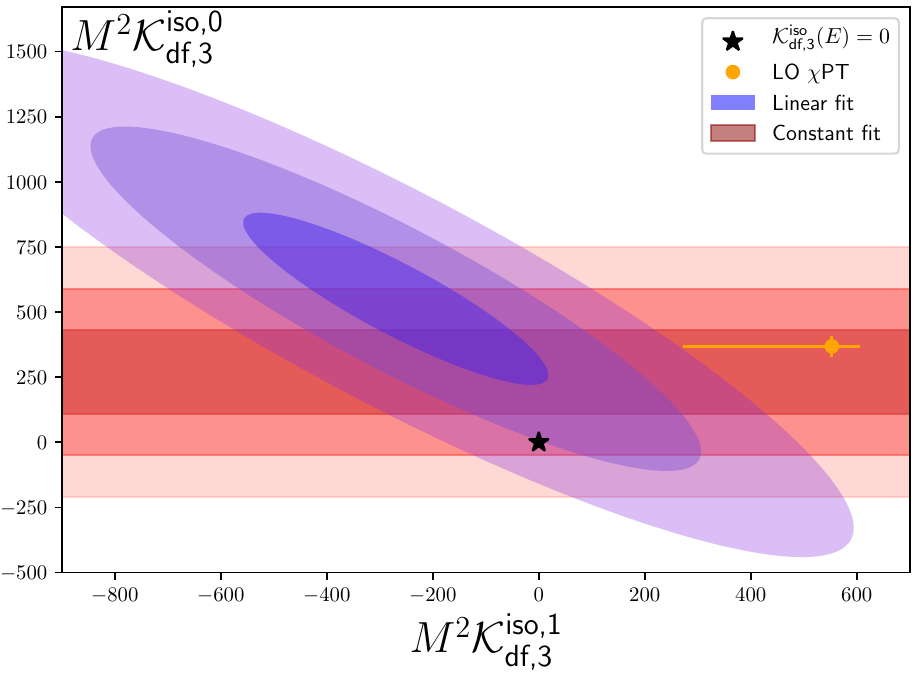}
\caption{One, two and  three-sigma confidence intervals for $M^2 \Kiso$ for the two different global fits
(4 and 5).
}
\label{fig:ConfKisoglobal}
\end{center}
\end{figure}

In Fig.~\ref{fig:ConfKisoglobal} we present a summary of the errors resulting from the global fits. 
We also include the value from LO \chpt, along with an estimate of the NLO corrections
obtained in Appendix~\ref{app:nonleading}, and quoted in Eq.~(\ref{eq:chpterrors}).
As can be seen, the constant term agrees well with the prediction, 
whereas the larger disagreement for the linear term is only of marginal significance
given the large uncertainty in the \chpt\ prediction.

One concern with our global fits is that we are using the forms for $\K_2$ and $\Kiso$ beyond their
radii of convergence. For $\Kiso$ we do not know the radius of convergence, but a reasonable
estimate is that one should use levels only with $|\Delta| < 1$. To check the importance of this issue,
we have repeated the global fits imposing $q^2/M^2 < 1$ and $\Delta < 1$, so that the fit includes
only five $2\pi^+$ and five $3\pi^+$ levels. We find fit parameters that are consistent
with those in Table~\ref{tab:global}, but with much larger errors. For example, the result from the
equivalent of fit 4 gives $M^2\Kisozero=610(350)$.

We close by commenting on sources of systematic errors. The results of Ref. \cite{\HH} are subject to discretization errors, but these are of $O(a^2)$, and likely small compared to the statistical errors from \cite{\HH}. The quantization condition itself neglects exponentially-suppressed corrections, but these are numerically small ($e^{-ML}\sim 1\%$) compared to our final statistical error. Errors from truncation of the threshold expansion for $\cK_2$ and $\Kdf$  are also present, but harder to estimate.

\section{Conclusions\label{sec:conc}}

We have presented statistical evidence for a nonzero $3\pi^+$ contact interaction, obtained by
analyzing the spectrum of three pion states in isosymmetric QCD with $M\approx 200\;$MeV
obtained in Ref.~\cite{\HH}.
This illustrates the utility of the three-particle quantization condition.
It also emphasizes the need for a relativistic formalism, since most of the
spectral levels used here are in the relativistic regime.
It gives an example where lattice methods can provide results for 
scattering quantities that are not directly accessible to experiment.

We expect that forthcoming generalizations to the formalism (to incorporate nondegenerate
particles with spin, etc.), combined with advances in the methods of lattice QCD 
(to allow the accurate determination of
the spectrum in an increasing array of systems), will allow generalization of the
present results to resonant three-particle systems in the next few years.

\section*{Acknowledgements}
We thank  Ra\'ul Brice\~no, Drew Hanlon, Max Hansen, Ben H\"orz and Julio Parra-Mart\'inez for discussions.
FRL acknowledges the support provided by the projects H2020-MSCA-ITN-2015//674896-ELUSIVES and H2020-MSCA-RISE-2015//690575-InvisiblesPlus and and FPA2017-85985-P. The work of FRL also received funding 
from the European Union Horizon 2020 research and innovation program 
under the Marie Sk{\l}odowska-Curie grant agreement No. 713673 and ``La Caixa'' Foundation (ID 100010434, LCF/BQ/IN17/11620044). The work of TDB and SRS is supported in part by the United States Department of Energy (USDOE) grant No. DE-SC0011637.


\appendix

\widetext
\begin{center}
\textbf{\large Supplementary material for the Letter:\\ $I=3$ three-pion scattering amplitude from lattice QCD}
\vspace{1cm}
\end{center}
\twocolumngrid
\makeatletter
\setcounter{equation}{0}
\setcounter{figure}{0}
\setcounter{table}{0}
\setcounter{page}{1}
\setcounter{equation}{0}
\setcounter{section}{0}
\makeatletter
\renewcommand{\theequation}{S\arabic{equation}}
\renewcommand{\thefigure}{S\arabic{figure}}
\renewcommand{\thesection}{S\arabic{section}}
\renewcommand{\thetable}{S\arabic{table}}

%
\section{Implementation of the QC in moving frames}
\label{app:moving}

Here we explain the essential features of the implementation of the RFT form of the quantization
condition, Eq.~(\ref{eq:QC}).
This has previously been carried out in the rest frame ($\vec P=0$), both keeping only
$s$-wave two-particle interactions and an isotropic $\Kdf$~\cite{\BHSnum},
and including $d$-wave two-particle interactions and the leading nonisotropic terms in $\Kdf$~\cite{\dwave}.
The generalization required here is to extend the $s$-wave plus isotropic $\Kdf$ approximation to
moving frames.

\subsection{$s$-wave approximation \label{app:movingswave}}

The matrices in Eq.~(\ref{eq:QC}) have indices $k, \ell, m$.
Here $k$ is shorthand for a finite-volume momentum, $\vec k= (2\pi/L)\vec n_k$,
which is labeled by an integer-valued vector $\vec n_k$.
This is the momentum of one of the three on-shell particles (denoted the spectator).
The other indices $\ell, m$ are the angular momentum quantum numbers of the remaining pair 
(called the interacting pair) in their center-of-mass frame.
The UV cutoff described below automatically cuts off $\vec k$, while formally $\ell$ runs over
all possible values. To obtain matrices of finite dimension we assume that $\cK_2$ is only nonzero
in the $s$-wave and that $\Kdf$ vanishes for $\ell>0$. It can then be shown that all solutions to
Eq.~(\ref{eq:QC}) that are sensitive to interactions are obtained by truncating all matrices to
have $\ell=m=0$.  In particular, $F_{3;p\ell' m',k\ell m}$ reduces to $F_{3; p k}^s$, where
we include the superscript as a reminder of restriction to $s$ waves.

The explicit form of $F_3^s$ is~\cite{\BHSnum}
\begin{equation}
L^3 F_3^s \equiv \frac{\wt F^s}{3} - \wt F^s \frac1{1/\wt \K_2^s + \wt F^s + \wt G^s} \, \wt F^s
\,,
\label{eq:F3s}
\end{equation}
where $\wt F^s$ and $\wt G^s$ are geometric matrices  
\begin{align}
\begin{split}
[\wt F^s]_{kp} & \equiv \frac{\delta_{kp}}{2} \frac{H(\vec k)}{2\omega_k}
\bigg [ \frac1{L^3}\sum_{\vec a} - \PV\int \frac{d^3 a}{(2\pi)^3} \bigg ] \times \\
& \ \ \ \ \ \ \ \ \ \ \ \ \ \frac{H_2(\vec a, \vec b)}{2 \omega_a   2\omega_b (E-\omega_k-\omega_a-\omega_b)}\,,
\label{eq:Fts}
\end{split}
\\ \begin{split}
[\wt G^s]_{kp} & \equiv \frac{H(\vec k) H(\vec p\,)}{ L^3 2  \omega_k 2\omega_p ( b^2 - m^2)}, \\
 b^\mu &= P^\mu -k^\mu -p^\mu
\end{split}
\label{eq:Gts}
\end{align}
and $\wt{\K}_2^s$ is given in terms of the $s$-wave phase shift by
\begin{align}
[1/{\wt {\K}_2^s}]_{kp} & \equiv  \delta_{kp} \big (1/{\wt \K_2^s(\vec k)} \big)\,,
\label{eq:K2tinv}
\\
{\wt \K_2^s(\vec k)} & \equiv
\frac{32 \pi \omega_k E_{2,k}^*}{
q_{2,k}^* \cot\delta_s(q_{2,k}^{*2}) + |q_{2,k}^*| [1-H(\vec k)]}\,,
\label{eq:K2t}
\end{align}
where
\begin{equation}
E_{2,k}^{*2} = (E-\omega_k)^2-(\vec P-\vec k)^2 \ \ {\rm and}\ \
q_{2,k}^{*2}= \frac{E_{2,k}^{*2}}4-m^2\,.
\label{eq:E2k}
\end{equation}
Other quantities appearing in these definitions are the on-shell energies, exemplified by
$\omega_k\equiv \sqrt{\vec k^2+m^2}$, the corresponding four-momenta, e.g.
$k^\mu=(\omega_k,\vec k)$, and the total four-momentum, $P^\mu=(E,\vec P)$.
Finally, the functions $H(\vec k)$ and $H_2(\vec a,\vec b)$ are UV cutoffs.
$H(\vec k)$ is a smooth function, cutting off the sum over $\vec k$ 
when $E_{2,k}^{*2}$ drops below zero,
and equaling unity for values of $\vec k$ such that the interacting pair lies above threshold.
We use the explicit form given in Refs.~\cite{\HSQCa,\BHSnum,\BHSQC,\dwave}, setting the
cutoff parameter to the value $\alpha_H=-1$.\footnote{%
The lower limit $E_{2,k}^{*2}=0$ (corresponding to $q_{2,k}^{*2}/m^2 = -1$) is not a fundamental
limit, but is related to the specific choice of boost used for below-threshold kinematics
in Refs.~\cite{\HSQCa,\BHSnum,\BHSQC,\dwave}. It is possible to choose a different form
of boost and allow $E_{2,k}^{*2}$ to become negative, and this option is currently under
investigation. 
}
For the cutoff function $H_2$ in $\wt F^s$ we use the ``KSS form'' given explicitly in 
Refs.~\cite{\BHSnum,\dwave}.

We observe that the only places where nonzero $\vec P$ enters above are into the definitions of
$E_{2,k}^*$, $q_{2,k}^*$, and $b^\mu$. Thus the numerical construction of the 
elements of the matrices is just as easy for moving frames as for rest frames.
The only complication arises when we project onto irreps, as discussed below.

Two-particle interactions enter through the K-matrix-like quantity $\wt \K_2^s$,
which we approximate by inserting the chosen parametrization of the phase shift, either Eq.~(\ref{eq:ERE}) or (\ref{eq:Adler}),
into Eq.~(\ref{eq:K2t}).
We note that, above threshold, where $H(\vec k)=1$, 
$2 \omega_k \wt \K_2^s(\vec k)$ is simply the standard two-particle K matrix.

To complete the quantization condition, we need the form of $\Kdf$ in the $s$-wave approximation.
At this stage, $\Kdf$ still depends on the spectator momenta $\vec k$ and $\vec p$.
However,  there is the additional constraint that the matrix form of $\Kdf$ is the restriction to
finite-volume momenta of an infinite-volume amplitude that is 
invariant both under Lorentz transformations and the exchange of both initial- and final-state particles.
We intuitively expect  that such a symmetric amplitude that is purely $s$-wave 
for any two-particle pair cannot depend on the spectator momentum.
One way to see that this is indeed the case is to use the threshold expansion developed in 
Ref.~\cite{\dwave}. At any order in the expansion parameter $\Delta$, one can show that the
 only terms that are purely $s$-wave are those that are isotropic.
Since nonisotropic terms first occur at $\cO(\Delta^2)$,
we work only at linear order in $\Delta$ in order to enforce $\ell_{\max}=0$
 within the context of the threshold expansion. Thus we use
\begin{equation}
\Kdf = \Kiso(\Delta) = \Kisozero + \Delta\, \Kisoone\,.
\label{eq:Kiso}
\end{equation}
This implies that, in matrix form, $\Kdf$ has the same entry in every element, and
is thus of rank 1.

As a result of these approximations, the QC reduces to
\begin{equation}
\det\left[ F_3^s(E,\vec P, L)^{-1}+ \Kiso(\Delta) \right] = 0
\,,
\label{eq:QCiso}
\end{equation}
where the dimension of the matrices is given by the number of finite-volume momenta
for which $H(\vec k)\ne 0$ for the given choice of $E$, $\vec P$,  and $L$.
The numerical problem is thus to find the eigenvalues of the matrix in Eq.~(\ref{eq:QCiso}) and
determine the energies at which they cross zero for the given choice
of the parameters in the $s$-wave phase shift and $\Kiso$.
This problem is greatly simplified in practice by block-diagonalizing the matrix, as we now explain.

\subsection{Block-diagonalization of the quantization condition}
\label{sec:irreps}

The energy levels of the finite-volume system fall into irreps of the relevant finite-volume
symmetry groups for various values of the total three-particle momentum $\vec{P}$.
For $\vec{P}=\vec{0}$, the symmetry group is the 48-dimensional cubic group $O_h$ (no double cover is needed since we are dealing with mesons).
The procedure for decomposing the QC in this case was first presented in Ref.~\cite{\dwave}, but the generalization to arbitrary $\vec{P}$ is new to this work and deserves explanation.
Unlike in Ref.~\cite{\dwave} where both $\ell_{\rm max}\in\{0,2\}$ were considered, 
here we focus only on the relevant case of $\ell_{\rm max}=0$.

For general $\vec{P}$, the finite-volume symmetry group is reduced to the little group $\text{LG}(\vec{P})$ of cubic group transformations $R\in O_h$ that leave $\vec{P}$ invariant:
\begin{align}
	\text{LG}(\vec P) \equiv \{ R\in O_h | R\vec{P} = \vec{P} \}.
\end{align}
We therefore seek to decompose the QC into irreps of $\text{LG}(\vec{P})$ instead of $O_h$, 
but otherwise the recipe used in Ref.~\cite{\dwave} is unchanged from that in Ref.~\cite{\dwave}.
The list of relevant little groups is shown in Table~\ref{tab:littlegroups}. 
\begin{table}[htp]
\begin{center}
\begin{tabular}{c|c|c|c}
$\vec d$ & $\text{LG}(\vec{P})$ & 2-pt.~irreps & 3-pt.~irreps  \\
\hline
$(0,0,0)$ 	& $O_h$ 	& $A_{1g}^+$, $E_g^+$	& $A_{1u}^-$, $E_u^-$ \\
$(0,0,1)$ 	& $C_{4v}$	& $A_1^+$, $B_1^+$		& $A_2^-$, $B_2^{-*}$ \\
$(1,1,0)$ 	& $C_{2v}$ 	& $A_1^{+*}$, $B_1^{+}$ 	& $A_2^-$ \\
$(1,1,1)$	& $C_{3v}$ 	& $A_1^+$, $E^+$		& $A_2^-$, $E^-$ \\
$(0,0,2)$ 	& $C_{4v}$	& $A_1^+$, $B_1^+$		& none
\end{tabular}
\end{center}
\caption{
Little group $\text{LG}(\vec{P})$ for each total momentum $\vec{P}=(2\pi/L)\vec{d}$ used in our fits, along with all irreps containing energy levels with $E_2^*\lesssim 4M$ or $E^*\lesssim 5M$.
We use the notation of Ref.~\cite{\HH} for irreps.
The asterisk indicates cases where the interacting energy lies slightly above the inelastic threshold,
although the free energy lies below.
}
\label{tab:littlegroups}
\end{table}

For fixed $(E,\vec{P},L)$, each matrix $M\in\{ \wt \K_2^s, \Kdf, \wt F^s, \wt G^s, F_3^s \}$ 
appearing in the QC is invariant under a common set of real unitary (i.e. orthogonal) 
transformations
%
$\{U(R)\}_{R\in \text{LG}(\vec{P})}$:
\begin{align}
	U(R)^TMU(R) &= M \quad \forall R\in \text{LG}(\vec{P})\,, \\
	U(R)_{p k} &= 
 \begin{cases} (-1)^{\Pi(R)}, &\vec{p}=R\vec{k} \\ 0, &\text{otherwise}. \end{cases}\, \label{eq:UR}
\end{align} 
Here $\Pi(R)$ is the parity of the transformation $R$, which is $+1$ if $R$ is a pure rotation
and $-1$ otherwise. This factor occurs because pions are pseudoscalars and leads to a
simple relabeling of irreps compared to those of scalars.
We note that the definition of $U(R)$ in Eq. \ref{eq:UR} differs from that in Ref.~\cite{\dwave} in three ways: it is sensitive to parity, it only includes $\ell=0$, and it has momentum indices transposed for notational convenience.

The transformation matrices $\{U(R)\}_{R\in \text{LG}(\vec{P})}$ furnish a (reducible) 
representation of $\text{LG}(\vec{P})$:
\begin{align}
	U(R_1R_2) &= U(R_1)U(R_2) \quad \forall R_1,R_2\in \text{LG}(\vec{P}), \\
	U(\mathbf{1_3}) &= \mathbf{1}_{k}.
\end{align}
This reducible representation can be decomposed into irreps $I$ of the little group $\text{LG}(\vec{P})$ through the use of projection matrices $P_I$:
\begin{align}
	P_I = \frac{d_I}{[\text{LG}(\vec{P})]}\sum_{R\in \text{LG}(\vec{P})} \chi_I(R)U(R)\,,
\end{align}
where $[\text{LG}(\vec{P})]$ is the dimension of the little group, $d_I$ is the dimension of $I$, 
and $\chi_I(R)$ is its character.
Lastly, we collect the eigenvectors of $P_I$ with nonzero (unit) eigenvalues into $P_{I,\text{sub}}$ to project QC matrices $M$ onto the lower-dimensional irrep subspace:
\begin{align}
	M_{I,\text{sub}} = \left( P_{I,\text{sub}} \right)^T M P_{I,\text{sub}}.
\end{align}
These projections partition the eigenvalues of $M$ into the various irreps of $\text{LG}(\vec{P})$,
so that we can study solutions to the QC irrep by irrep.

The isotropic nature of $\Kiso$ implies that it contributes only to the most symmetric irreps, modulo
the presence of parity. In particular, for $\vec d=0$ it contributes only to the $A_{1u}^-$ irrep,
while for $\vec d^2=\{1,2,3\}$ it contributes only to the $A_2^-$ irrep of the respective little groups (see Table~\ref{tab:littlegroups}).
For these irreps, one can use the arguments presented in Refs.~\cite{\HSQCa}
and extended in Ref.~\cite{\dwave} to reduce the QC to a one-dimensional algebraic relation,
referred to as the isotropic approximation to the QC.
This is not necessary, however, as one can instead simply project onto these irreps as described
above. In practice we have used both methods and checked that they agree.

For all other irreps that arise in the elastic portion of the three-particle spectrum, $\Kiso$ does not
contribute. This does not mean, however, that there is no shift of the energies from their noninteracting
values, as the two-particle interactions do indeed lead to energy shifts. This point was not appreciated
in previous work, where it was claimed that there would be no energy shifts in such irreps~\cite{\dwave}. 
The presence of a shift can be understood intuitively in the case of 
the $E_u^-$ irrep for $\vec P=0$.
Here the lowest total angular momentum contained is $J=2$. Thus the interacting pair can be in
an $s$-wave, and thus affected by the $s$-wave two-particle interaction, with the total $J=2$ being
obtained by having the spectator in a $d$-wave relative to the pair.
These irreps (of which there turn out to be three --- see Table~\ref{tab:nontrivialirreps}) provide
an additional constraint on the two-particle amplitude.

The generalization of these considerations to include $d$-waves in moving frames is straightforward,
but beyond the scope of this work.

\section{Non-leading effects in $\Kdf$}
\label{app:nonleading}

We now provide further justification for the results
\begin{align}
M^2 \Kdf &= M^2 \Mdf (1 + \cdots ) 
\label{eq:KtoM2}
\\
&= \frac{M^4}{F^4} (18 + 27 \Delta + \cdots)
\label{eq:Kdfres2}
\end{align}
discussed in the main text [see Eqs.~(\ref{eq:Mdfres}) and (\ref{eq:KtoM})],
and estimate the size of the NLO corrections indicated by the ellipses.

\subsection{Derivation of Eq.~(\ref{eq:KtoM2})}

We first discuss the derivation of, and corrections to, Eq.~(\ref{eq:KtoM2}).
This result provides
the second step in the two-part relation between
the  three-particle spectrum and the physical three-particle scattering amplitude.
The equations governing this step were derived in Ref.~\cite{\HSQCb}.
In the case that only $s$-wave two-particle channels interact, and $\Kdf$ is isotropic,
$\Mdf$ is also restricted to the $s$-wave, but is not, in general, isotropic.
Specifically, it is given by~\cite{\HSQCb,\BHSnum}
\begin{equation}
\Mdf(p,k) = \cS\left\{ \frac{\cL(k)\cL(p)}{1/\Kiso+ F_3^\infty} \right\}\,,
\label{eq:KtoM}
\end{equation}
where $\cS$ denotes symmetrization over momentum assignments.
The other quantities in Eq.~(\ref{eq:KtoM}) are
\begin{align}
\cL(k) &= \frac13 \!-\! 2\omega(k)\cM_2^s(k) \wt \rho(k) \!-\! \int_{\vec s} \cD^{(u,u)}(k,s)\wt\rho(s)\,,
\label{eq:ell}
\\
\wt\rho(k) &= \frac{H(k)}{2\omega(k)} \rho(k)\,,\qquad
\rho(k) = \frac{|q_{2,k}^*|}{16 \pi E_{2,k}^*}\,,
\label{eq:rho}
\end{align}
where we have given only the subthreshold form of $\rho$.
In the previous equations, all  integrals are three-dimensional
\begin{equation}
\int_{\vec s} = \int \frac{d^3s}{(2\pi)^3}
\,,
\end{equation}
despite the fact that the integrands depend only on the magnitudes of the momenta.

Thus to determine $\cL(k)$, we need $\cD^{(u,u)}$, the asymmetric form of $\cD$ in Eq.~\eqref{eq:D} of the main text, which is given by solving
\begin{multline}
\cD^{(u,u)}(k,p) = -\cM_2^s(k) G^\infty(k,p) \cM_2^s(p) \\
-\int_{\vec s} \frac1{2\omega(s)} \cM_2^s(k) G^\infty(k,s) \cD^{(u,u)}(s,p)\,,
\label{eq:Duu}
\end{multline}
where we use the relativistic form of $G^\infty$~\cite{\BHSQC}
\begin{equation}
G^\infty(k,p) = \frac{H(k) H(p)}{(P-k-p)^2-m^2+i\epsilon}
\,.
\label{eq:Ginfty}
\end{equation}
Finally, $F_3^\infty$ is given by
\begin{equation}
F_3^\infty = \int_{\vec k} \wt\rho(k) \cL(k)\,.
\label{eq:F3infty}
\end{equation}

The above equations depend on the physical two-particle $s$-wave scattering amplitude, $\cM_2^s$,
which is  related to $\cK_2^s$ by
\begin{align}
\frac1{\cM_2^s(k)} &= \frac1{\cK_2^s(k)} + \rho(k)\,,
\end{align}
with $\cK_2^s$ given by
\begin{align}
\frac1{\cK_2^s(k)} &= \frac1{16\pi E_{2,k}^*} q_{2,k}^* \cot \delta(q_{2,k}^*)
\,.
\end{align}
The phase shift is given in turn by the parametrization of Eq.~\ref{eq:Adler}.

We can now explain in detail why $\Mdf=\Kdf$ at LO in \chpt, i.e. why Eq.~(\ref{eq:KtoM2}) is valid.
The power-counting parameter is $\epsilon \sim M^2/F^2 \sim k^2/F^2$, in terms of
which $\cM_2^s\sim \epsilon$, $\tilde \rho \sim 1$, and $G^\infty \sim 1/\epsilon$, implying that
that $\cD^{(u,u)}\sim \epsilon$. 
It follows that $\cL= 1/3 + \cO(\epsilon)$, with the determination of the $\cO(\epsilon)$ terms
requiring the full solution to the integral equation for $\cD^{(u,u)}$.
In addition, we see that $F_3^\infty \sim \epsilon^0$.
Since we know from Eq.~(\ref{eq:Mdfres}) that $\Mdf \sim \epsilon^2$, it then follows from 
Eq.~(\ref{eq:KtoM}) that also $\Kiso\sim \epsilon^2$, as this is the only way to match
powers of $\epsilon$ on the two sides. Thus the $F_3^\infty$ term in the denominator is
actually of NNLO relative to the dominant $1/\Kiso\sim 1/\epsilon^2$ contribution.
In summary, at LO we can set $\cL\to 1/3$ and drop $F_3^\infty$. Symmetrization leads to a factor
of 9 that cancels the $(1/3)^2$, leading to $\Mdf=\Kiso$.

To complete the discussion, we note that the restriction to the $s$-wave, isotropic approximation
is also consistent with \chpt. In particular, the $d$-wave amplitude 
and the nonisotropic part of $\Kdf$ appear first at NLO, since both require an additional
factor of $k^2$ relative to the corresponding LO amplitudes. Thus, at LO, $\cM_2=\cM_2^s$
and $\Kdf=\Kiso$. Therfore the result just derived, $\Mdf=\Kiso$ at LO, is equivalent to
$\Mdf=\Kdf$. A further check on this is provided by the fact that the LO result for $\Mdf$ calculated
explicitly in the main text is isotropic.

\subsection{Higher order corrections in relation between $\Mdf$ and $\Kdf$}

We now turn to an estimate of higher-order corrections to Eq.~(\ref{eq:KtoM2}).
This is provided by solving the equations above and determining the quantities
$3\cL(k)-1$, which is of NLO, and $\Kiso F_3^\infty$, which is of NNLO.
This is not a complete calculation of higher-order corrections, since, as already noted,
$d$-wave amplitudes and nonisotropic contributions to $\Kdf$ also appear at NLO.
A further approximation is that we solve the equations 
only below the three-particle threshold, so that the pole
prescription in $G^\infty$ is not needed. This is sufficient to determine the order of magnitude
of the corrections. Methods for solving the equations above threshold have not yet been developed.
For convenience, we set $\vec P=0$, although this makes no difference to the final result, which
is relativistically invariant.
This choice implies that, in this subsection and the next,
$E$ and $E^*$ are equal, and we use the former for brevity.

Following Ref.~\cite{\BHSnum},
the equations are solved by discretizing the momenta as though the system was
in a periodic box of size $L$, leading to $\vec k=(2\pi/L)\vec n$ with $\vec n\in\mathbb{Z}^3$.
One then has to invert large matrix equations, which is straightforward.
This also allows us to reuse much of the setup needed for the QC itself, since the equations above
are obtained by taking the $L\to\infty$ limit of various objects that appear in the QC.
We stress, however, that here we are using the finite-volume simply as a device for solving the integral
equations and that $L$ here is not related to the volume in which actual lattice QCD simulations are done.

The function $\cL(k)$ is given by Eq.~(A14) of Ref.~\cite{\BHSnum}:
\begin{equation}
\begin{split}
&\cL(k) = \lim_{L\to\infty} \cL(k;L)
\\
= \lim_{L\to\infty} 
\Bigg\{ \frac13 -& \sum_{\vec p} \left[
\frac1{1/(2 \omega \cM_2^s) + \wt G^s}\, \wt \rho \right]_{kp} \Bigg\}\,,
\end{split}
\end{equation}
where the indices $k$ and $p$ are drawn from the finite-volume set,
 $\omega$, $\cM_2^s$ and $\wt\rho$ are simply diagonal matrices
 containing entries of the corresponding infinite-volume quantities evaluated at
 finite-volume momenta,
and the matrix $\wt G^s$ is 
\begin{equation}
\left[\wt G^s\right]_{kp} = \frac1{L^3}\frac1{2\omega_k} G^\infty(k,p) \frac1{2\omega_p}\,.
\end{equation}
$F_3^\infty$ is given by the discretized form of Eq.~(\ref{eq:F3infty}),
\begin{equation}
F_3^\infty = \lim_{L\to\infty} \frac1{L^3} \sum_{\vec k} \wt\rho(k) \cL(k,L)\,.
\end{equation}
We find that the convergence in $L$ is quite rapid, with $ML \gtrsim 30$ enough to obtain the
form of the solution to sufficient accuracy.

\begin{figure}[t]
\begin{center}
\includegraphics[width=.47\textwidth]{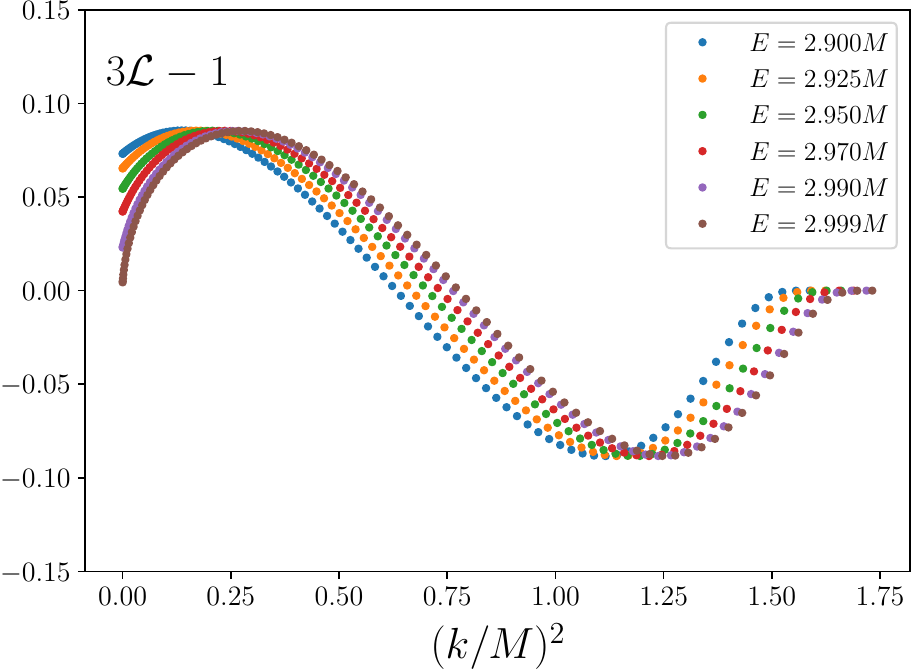}
\caption{Dependence of $3\cL(k)-1$ on $(k/M)^2$ for $ML=60$ and 
different values of the rest-frame energy,  $E/M$. 
This quantity indicates the size of the NLO corrections in the relationship between $\Kdf$ and $\Mdf$.}
\label{fig:3Lminus1}
\end{center}
\end{figure}

\begin{figure}[t]
\begin{center}
\includegraphics[width=.47\textwidth]{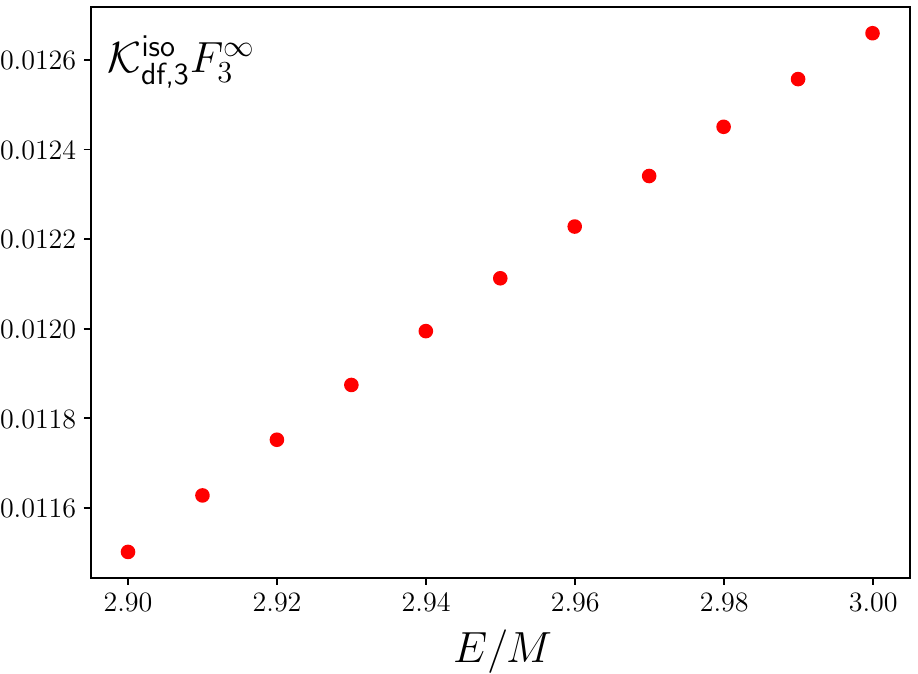}
\caption{$\Kiso F_3^\infty$ as a function of the energy, $E/M$. We set $M^2 \Kiso=550$.
This quantity indicates the size of NNLO corrections in the relationship between $\Kdf$ and $\Mdf$.
}
\label{fig:F3inf}
\end{center}
\end{figure}

{In order to display results, we choose fit 1 from Table~\ref{tab:resAdler} for the phase shift
that determines $\cM_2^s$.
In Fig.~\ref{fig:3Lminus1}, we plot the NLO quantity $3\cL(k)-1$ as a function of $k^2/M^2$ 
for $2.9\le E/M < 3.0$. 
This quantity necessarily vanishes at the maximum value of $k$, set by the value in which the cutoff function vanishes, $H(k)=0$. 
We observe that the correction has a maximum of about 0.09, and is smaller for $k$ near zero. Indeed, extrapolating to $E=3M$, we find
$3\cL(0)-1 \approx -0.005$ at threshold.
The NNLO quantity  $\Kiso F_3^\infty$ is shown in Fig.~\ref{fig:F3inf}, using
$M^2\Kiso=550$, the value obtained in the first global fit in Table~\ref{tab:global}. 
The small result, at the percent level, is consistent with this being a higher-order effect.
}

\subsection{Estimating higher order corrections to $\Kdf$}

Here we estimate the corrections in Eq.~(\ref{eq:Kdfres2}) that are indicated by the ellipses.
First we consider the value of $\Kdf$ at threshold, where $\Delta=0$. Corrections arise both
in the relation between $\Kdf$ and $\Mdf$, and in the \chpt\ result for $\Mdf$ itself.
The results just obtained for $3 \cL(0)-1$ and $\Kiso F_3^\infty$ at threshold
imply few percent relative corrections in the $\Kdf$ to $\Mdf$ relation.
Higher order corrections in the result for $\Mdf$ are expected to be of generic relative size
$M^2/(4\pi F)^2 \approx M a_0/\pi \approx 0.03$. Assuming constants of $\sim 3$ multiplying these generic
corrections, we estimate them conservatively to be no larger than 10\%. These generic corrections
thus dominate the error estimate at threshold.

Next we consider the corrections to the linear term in $\Delta$ in Eq.~(\ref{eq:Kdfres2}).
We expect the generic corrections to be of similar relative magnitude as at threshold, i.e. $\lesssim 10\%$.
The corrections to the $\Kdf$ to $\Mdf$ relation can, however, be larger.
We focus on the dominant contribution, that from $\cL(k)\cL(p)$ in Eq.~(\ref{eq:KtoM}).
The momentum dependence of $\cL(k)$ near threshold implies that a constant $\Kiso$ will lead to
a $\Delta$ dependence in $\Mdf$, and {\em vice versa}. In particular, if we fix $\Kiso$ to
a constant, and calculate the derivative
\begin{equation}
c = \frac1{\Kiso} \frac{d \Mdf}{d \Delta}\Bigg|_{\Kiso,\Delta=0}\,,
\end{equation}
then we have, for small $\Delta$, and ignoring the generic \chpt\ corrections,
\begin{equation}
M^2 \Kdf = \frac{M^4}{F^4} \left[18 + (27 -18 c) \Delta + \cO(\Delta^2) \right]\,.
\label{eq:Kdfnew}
\end{equation}
In words, the constant feeds down a correction to the linear term.

To estimate $c$, we use the results of Fig.~\ref{fig:3Lminus1}. These are calculated for
$E < 3M$, corresponding to $\Delta< 0$. Recalling that $\Mdf$ and $\Kdf$
are on-shell amplitudes, we observe that to obtain $\Delta<0$ we require $k^2 < 0$.
For example, a configuration $p_1^\mu=(M,\vec 0)$, 
$p_2^\mu= (0,i M,0,0)$ and $p_3^\mu=(0,-iM,0,0)$ has all particles on shell, $E^2=M^2$, 
and thus $\Delta=-8/9$. Taking each of the particles in turn as the spectator, the
values of $k^2/M^2$ are $0$, $-1$ and $-1$, respectively (remembering that $k^2\equiv \vec k^2$).
Averaging over the choices of spectator, we find $\langle k^2/M^2 \rangle=-2/3$ for $\Delta=-8/9$.
In principle, one should do an average over all allowed momentum configurations, but our simple
example gives a rough relation between $\Delta$ and $k^2$, namely
\begin{equation}
\frac{d \Delta}{d (k^2/M^2)}\Bigg|_{k^2=0} \approx 4/3 \,.
\end{equation}
The final step is to use Fig.~\ref{fig:3Lminus1} to estimate
\begin{equation}
\frac{d\, 3\cL(k)}{d (k^2/M^2)}\Bigg|_{k^2\approx 0, E\approx 3M} \approx \frac12\,.
\end{equation}
This is a crude estimate, given that the slope depends on $E$.
Nevertheless, using these results, and the fact that there are two factors of $\cL$ in Eq.~(\ref{eq:KtoM}),
we arrive at the estimate
\begin{equation}
c \approx 2 
\frac{\frac{d\, 3\cL(k)}{d (k^2/M^2)}\bigg|_{k^2\approx 0, E\approx 3M} }
{\frac{d \Delta}{d (k^2/M^2)}\bigg|_{k^2=0}}
\approx 0.75
\,.
\end{equation}
Inserting this into Eq.~(\ref{eq:Kdfnew}) we find that the term linear in $\Delta$ is reduced by
about 50\% by this correction. We treat this as an asymmetric error, 
since the sign of the effect is unambiguous.
We do not shift the central value, as the error estimate is itself uncertain.

In summary the \chpt\ prediction for $\Kdf$ becomes
\begin{equation}
\begin{split}
M^2 \Kisozero &= \frac{M^4}{F^4} 18 (1 \pm 10\%) = 360 \pm 36 \,, 
\\
M^2 \Kisoone &= \frac{M^4}{F^4} 27 (1 {}^{+10\%}_{- 51\%} ) = 540\, {}^{+54}_{-275} \,,
\end{split}
\label{eq:chpterrors}
\end{equation}
where numerical values are obtained using $M a_0= 0.089$ from fit 1.

\section{Further details on fits}
\label{app:fits}

In this section we  provide a more detailed explanation of our fitting procedures,
and further details of the results of the fits.

\subsection{General fitting procedure}

We determine $\K_2$ and $\Kdf$ by fitting solutions to the two-
 and three-particle QCs to the energy levels provided in Ref.~\cite{\HH}, 
 which were computed on the CLS D200 $N_f=2+1$ ensemble,
which has pion mass $M  \sim 200 \text{ MeV}$, lattice size $64^3\times 128$
and inverse lattice spacing $1/a \approx 3.1\;$GeV~\cite{Bruno:2014jqa,Bruno:2016plf}. 
These parameters imply that $M L\approx 4.2$, which is large enough that we 
expect neglected exponentially-small corrections are at the percent level.

The three-particle QC, Eq.~\eqref{eq:QC}, has been discussed above.
The two-particle QC for states that couple to $J=0$ can be written as
\begin{equation}
q \cot \delta_0(q) = \frac{2}{\gamma L \sqrt{\pi}}  Z_{00}(q^2, \vec{d}), \label{eq:Luscher}
\end{equation}
where $Z_{00}$ is the standard L\"uscher Zeta function, $\vec P = (2\pi/L)\vec d$ is the total two-particle momentum, $\gamma$ is the boost factor to the center-of-mass frame, and $q^2=E_2^{*2}/4-M^2$.
As discussed in the main text, we consider two parametrization schemes for $\delta_0$: 
the standard ERE of Eq.~\eqref{eq:ERE} and the Adler-zero form of Eq.~\eqref{eq:Adler}.
The parameters in the two schemes can be related by expanding the Adler-zero form about threshold:
\begin{align}
M a_0 &= - \frac1{B_0} \frac{2 M^2-z_2^2}{M^2} \xrightarrow{z_2\to M} - \frac1{B_0}\,,
\label{eq:a0Adler}
\\
M^2 r a_0 &= - \frac{2 B_1}{B_0}  + \frac{2M^2+z_2^2}{2M^2-z_2^2}
\xrightarrow{z_2\to M}  - \frac{2 B_1}{B_0}  + 3\,.
\label{eq:ra0Adler}
\end{align}

Once we choose a parametrization scheme for $\K_2$ (and $\Kdf$ for three-particle energies), we fit the parameters by minimizing the following $\chi^2$ function \cite{Dudek:2012gj}:
\begin{align}
	\chi^2 = \sum_{i,j}  \left(  E^*_i - (E^*_i)^{sol} \right)  C^{-1}_{ij}   \left( E^*_j- (E^*_j)^{sol} \right), \label{eq:chi2}
\end{align}
where $\{E^*_i\}$ are the center-of-mass energy levels of Ref.~\cite{\HH} with covariance matrix $C$, and $\{(E^*_i)^{sol}\}$ are the solutions to the appropriate QC(s) for a particular set of parameters.
To estimate the statistical uncertainties of our fit parameters, we use the individual bootstrap samples provided by Ref.~\cite{\HH} to perform multiple fits for each scheme.
We note that the correlation matrix $C$ is taken to be the same for all bootstrap samples.

\subsection{Results of additional fits to $\cK_2$}

We have fit the Adler zero form (\ref{eq:Adler}) to the restricted data set of the five two-particle 
levels that lie inside the formal radius of convergence of the expansion, $|q^2|/M^2 < 1$.
The results are given in Table~\ref{tab:resAdler2}, 
which should be compared to Table~\ref{tab:resAdler}.
The main conclusion is that the fits yield compatible parameters, providing a consistency check
on the results obtained in the main text.
The errors here are larger, as expected, and, indeed, very large in fits 7 and 8, where there
are insufficient data points to determine the three parameters.

\begin{table*}[t]
\centering
\begin{tabular}{cccccccc}
Fit                         & $B_0$ & $ B_1$ & $B_2$ & $z^2_2/M^2$ & $\chi^2/$dof & $M a_0$ & $M^2 r a_0$ \\ \hline\hline
6 &   -10.9(1.0) &  -2.5(2.3)         &   ---    &    1 (fixed)     &    2.89/(5-2)  &\  0.092(8) \  &  2.5(4)  
\\ \cline{1-8} 
 7  &  -11.0(1.1) & -2(5)&   -1(6)     &  1 (fixed)      & 2.86/(5-3)    & \ 0.091(9) \ & 2.7(9)    
 \\ \cline{1-8} 
  8  &   -11(8) &  -3(7)   &--- &1.0(8)      &     2.89/(5-3)  &\  0.091(9) \ &  2.6(1.9) 
\end{tabular}
\caption{Fits of the two-particle spectrum to the Adler-zero form of $q\cot\delta_0$, Eq.~(\ref{eq:Adler}),
considering only levels for which $q^2/M^2 < 1$.
}
\label{tab:resAdler2}
\end{table*}

We have also repeated the Adler-zero fits to all levels in the elastic region using the ERE form, 
Eq.~(\ref{eq:ERE}).
Although this is not justified theoretically (since the radius of convergence is $|q^2|/M^2 \lesssim 1/2$),
it provides a comparison with a standard form that has been used in previous lattice calculations of the $I=2$ two-pion amplitude. The results are shown in Table~\ref{tab:resERE}.
The quality of the fits is poor, and the results for $M^2 r a_0$ are in strong disagreement
with the LO \chpt\ prediction of 3. This provides additional support to the theoretical arguments
favoring the use of the Adler-zero fit form.

\begin{table}[t]
\centering
\begin{tabular}{cccccc}
Fit                         & $M a_0$ & $M r$ & $P$ & $\chi^2$/dof  & $M^2 r a_0$\\ \hline\hline
9 &  0.114(6) & 2.15(29)         &  ---      &          29.81/(11-2)   &  0.25(3)   \\ \cline{1-6} 
 10                   &  0.106(8) & 5.1(1.3) & -0.0030(14)    &          23.54/(11-3)  & 0.54(11)     
\end{tabular}
\caption{Fits of the two-particle spectrum to the ERE form of $q\cot\delta_0$, Eq.~(\ref{eq:ERE}).
All levels up to (and slightly beyond) the elastic threshold at $q^2/M^2=3$ are used, as in
fits 1-3 in Table~\ref{tab:resAdler}.
}
\label{tab:resERE}
\end{table}

\subsection{Determining the $d$-wave scattering length}

To study two-particle $d$-wave interactions, we analyze the energy levels 
of Ref.~\cite{\HH} that lie in irreps that do not couple to $s$-wave interactions. 
For each such level, Table~\ref{tab:dwave} shows the comparison of the determined energy to
the corresponding free energy.
All energy shifts are small and positive, suggesting a very mildly repulsive $d$-wave interaction.
To quantify this interaction, we use the ERE:
\begin{align}
	q\cot\delta_2(q) = \frac{1}{q^4}\left[ \frac{1}{(a_2)^5} + \cO(q^2) \right]\,. \label{eq:a2}
\end{align}
We then extract the $d$-wave scattering length $a_2$ using the $d$-wave form of the two-particle QC
 (see, e.g., Refs.~\cite{Luu:2011ep,Gockeler:2012yj}), yielding
\begin{align}
\begin{split}
(M a_2)^5 &= 0.0006(2),  \\ \chi^2/\text{dof}=3.3&/(5-1)=0.83. \label{eq:dwave}
\end{split}
\end{align}
This result is nonzero with $3\sigma$ significance.
It is, however, numerically small, suggesting that we can neglect it in our fits to the three-particle levels.

\begin{table}[htp]
\begin{tabular}{c|c|c|c}
irrep &                $E_2^*/M({\rm free})$ &      $E_2^*/M({\rm  interacting})$~\cite{\HH} &     difference \\
\hline
$E_2^-(0)$ &   3.621(13)  &  3.624(13)  & 0.003(3)\\
$B_1^-(1)$ &   3.885(14) &  3.889(15)  & 0.004(4)\\
$B_1^-(2)$ &   4.086(17) &  4.091(16)  & 0.005(2)\\
$E^-(3)$     &   3.246(10)  &     3.246(10)  & 0.000(2)\\
$B_1^-(4)$ &   3.621(13) &    3.628(13)  & 0.006(2)
\end{tabular}
\caption{Comparison of free and interacting spectra (the latter from Ref.~\cite{\HH}) for two-particle states in irreps that do not couple to $\ell=0$. 
The number in parentheses for each of the irreps gives $\vec d^2$ and thus specifies the frame.
}
\label{tab:dwave}
\end{table}

To study this further, we examine the systematic error induced in the three-particle spectrum by neglecting the $d$-wave scattering length. For this we study the effect of $a_2$ on the three-particle energy levels in the rest frame, 
where we have previously implemented the three-particle QC including both $s$- and
$d$-wave effects~\cite{\dwave}.
Taking $\cK_2$ from the first fit of Table \ref{tab:resAdler},
and $a_2$ from Eq.~(\ref{eq:dwave}), we find the
results shown in Table \ref{tab:dwave3part}.
We see that $d$-wave effects are completely negligible in the $A_1^-$ irrep, and less than a third of the statistical error in the $E^-$ irrep.
We therefore expect that, for current precision, $d$-wave effects can safely be ignored.

\begin{table}[h!]
\begin{tabular}{c|c|c}
irrep &           $E^*/M$~\cite{\HH} &    $\delta E^*(a_2)/M$ \\
\hline
$A_1^-(0)$ &     4.780(17)  & 0.0004(2) \\
$E^-(0)$ &     4.691(15) &  0.005(2) \\
\end{tabular}
\caption{Effect of $d$-wave interactions on the three-particle energy levels in the rest-frame. 
Here $E^*$ is the center-of-mass energy of the level,
while $\delta E^*( a_2) = E^*(a_2) - E^*(0)$ is the shift in this energy upon inclusion of the nonzero 
$a_2$ given in Eq.~(\ref{eq:dwave}). We have fixed $M^2\Kiso=500$, but the results are
insensitive to this value. Other notation as in Table~\ref{tab:dwave}.}
\label{tab:dwave3part}
\end{table}

\subsection{Fitting $\Kiso$ using method 1}

Here we provide more details regarding our fits to determine $\Kiso$ using method 1, which was
described briefly in the main text.

Within each bootstrap sample, we 
(a) 
fit the simplest Adler-zero form for ${\K}_2^s$ (fit 1---see Table \ref{tab:resAdler}),
to the eleven two-particle levels that are sensitive to $s$-wave interactions and lie
below (or slightly above) the inelastic threshold at $E_2^*=4M$;
and (b)
determine the values of $\Kiso$ that, when inserted in the QC, give the energies of each of the 
eight three-particle energy levels which are
sensitive to $\Kiso$ and lie below (or slightly above) the inelastic threshold at $E^*=5M$.
Averaging over bootstrap samples in the standard way, we obtain the average values for
each of the eight $\Kiso$ values, as well as the correlation matrix between them.
Using this correlation matrix, 
we then do a standard fit to the results for these eight levels, either using a constant or a linear
form in $\Delta$.

Fitting to a constant yields
\begin{align}
\begin{split}
M^2\Kiso &= 560(270), \\ \chi^2/\text{dof } &= 8.5/(8-1) = 1.21,
\end{split}
\end{align}
while a linear parametrization gives
\begin{align}
\begin{split}
M^2\Kiso &= 140(430)+ 570(500)  \Delta , \\ \chi^2/\text{dof } &= 7.7/(8-2) = 1.28. \label{eq:fitlin1}
\end{split}
\end{align}
The constant fit points towards a $2\sigma$ significance on $\Kiso$.
For the linear fit we note that the errors highly correlated, and thus even though each parameter is compatible with zero, the point where both vanish and $\Kdf(E)=0$ is also excluded by $2\sigma$.
These results are shown in Fig.~\ref{fig:Kiso} of the main text.

Finally, as a consistency check, we use the QC to predict the energies
for those irreps that are not affected by $\Kiso$, with results shown in Table~\ref{tab:nontrivialirreps}.
We find that the predicted values lie very close to the measured values.
This indicates that our restriction to $s$-waves, and our parametrization of ${\K}_2^s$,
are sufficient given present precision.
We therefore include these energy levels in our global fits.

\begin{table}[h!]
\begin{tabular}{c|c|c}
irrep &           $E^*/M$~\cite{\HH} &    prediction \\
\hline
$E^-(0)$ &     4.691(15)  &  4.685(14)\\
$B_2^-(1)$ &    5.008(17) &  5.007(16) \\
$E^-(3)$ &     4.528(14) &  4.529(13) \\
\end{tabular}
\caption{Prediction for the three-particle energy levels in irreps that are insensitive to $\Kiso$. 
Notation as in Table~\ref{tab:dwave}.}
\label{tab:nontrivialirreps}
\end{table}

\subsection{Correlation matrix for global fits}

Here we collect the covariance matrices for the global fits in Table~\ref{tab:global}. 
We write these as $C = D R D$, with $D$ the diagonal matrix 
containing the standard errors in the parameters. Our results are
\begin{align}
\text{ Fit 4: } &D = \text{diag }\left(0.7,0.3,160\right),  \\ 
&R = \begin{pmatrix}
1 & -0.67 & 0.23   \\
-0.67 & 1 & 0.24   \\
0.23 & 0.24 & 1  \\
\end{pmatrix}, \\
\text{ Fit 5: } &D = \text{diag }\left(0.7,0.3,330,290\right),  \\ &R = \begin{pmatrix}
1 & -0.63 & 0.22 & -0.10  \\
-0.63 & 1 & -0.11 & 0.25  \\
0.22 & -0.11 & 1 & -0.89  \\
-0.10 & 0.25 & -0.89 & 1  \\
\end{pmatrix}, \hspace{-0.1cm}
\end{align}
where the matrix indices are ordered as $(B_0, B_1,M^2\Kisozero )$ and $(B_0, B_1,M^2\Kisozero ,M^2\Kisoone )$, respectively. 
As can be seen, the correlation is large within the two- and three-particle sector, 
and smaller between the two different sectors.

\subsection{Two- and three-pion spectrum}

\begin{figure}[h]
   \centering
   \subfloat[\ $2\pi^+$ spectrum. \label{fig:spec2}]%
             {\includegraphics[width=0.49\textwidth,clip]{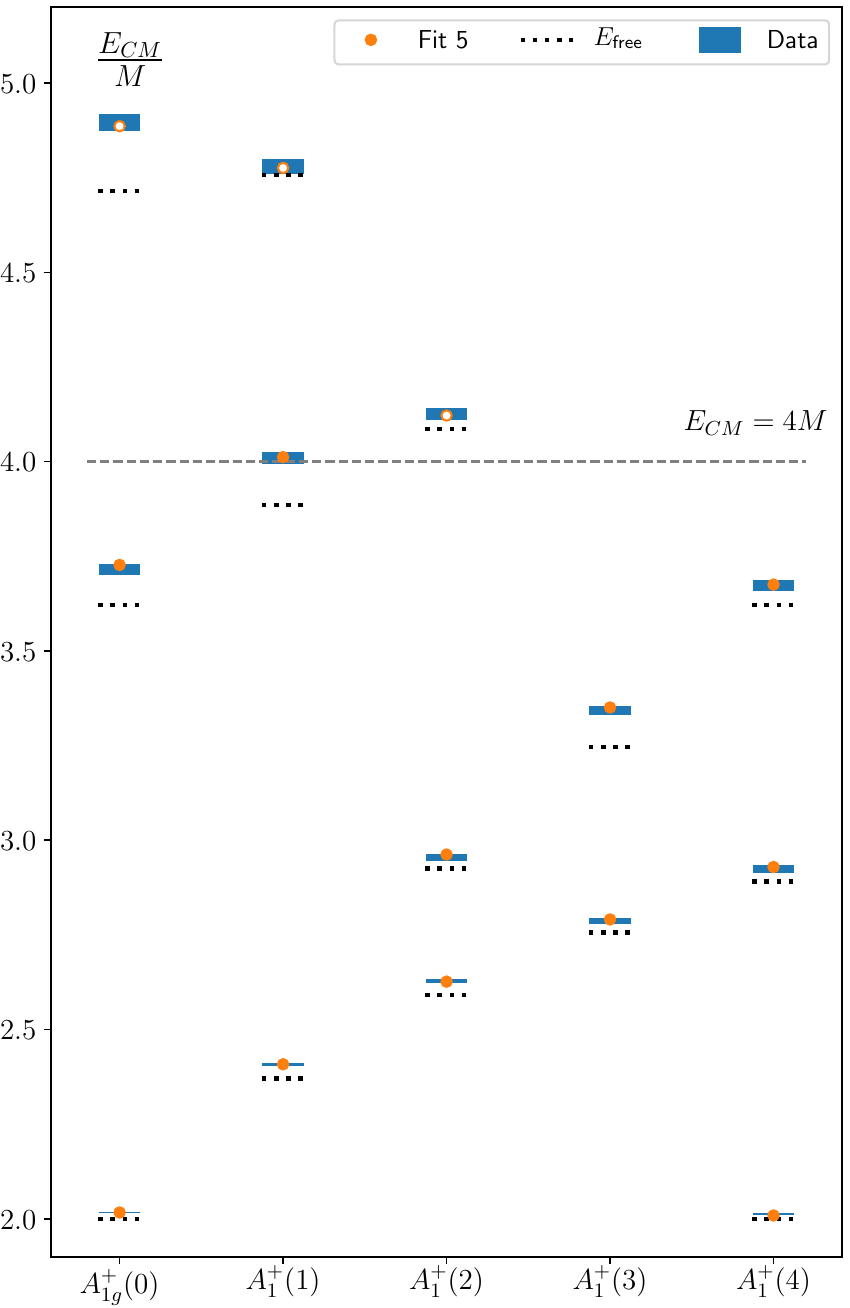}}\hfill
   \subfloat[ \ $3\pi^+$ spectrum. \label{fig:spec3} ]%
             {\includegraphics[width=0.488\textwidth,clip]{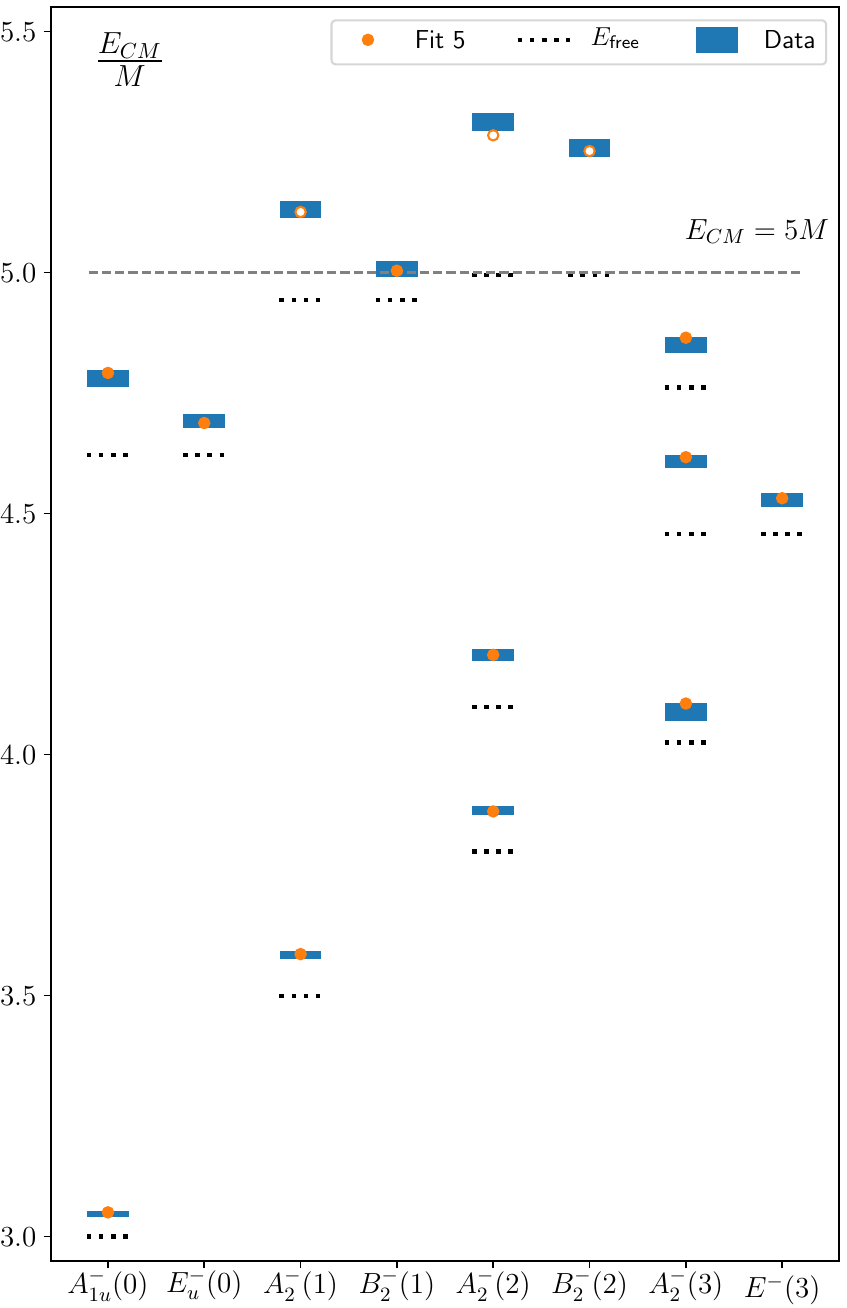}}
   \caption{ Two- and three-pion spectra from Ref. \cite{\HH} (blue) compared to the predictions from the global fit 5 (orange). Hollow orange points above the inelastic thresholds have not been included in the fit, but are shown for comparison. Dashed lines show the non-interacting energy levels. }
   \label{fig:spectra}
\end{figure}

To conclude, we provide a comparison of the data to the predicted two- and three-pion spectra from the quantization conditions. For this, we use the best parameters from fit 5 described in the main text (see Table \ref{tab:global}). The results are displayed in Fig. \ref{fig:spectra}. We also include the predictions from the QC above the inelastic thresholds--- $E_{CM}=4M$ and $E_{CM}=5M$ for the two- and three-particle QC, respectively. As can be seen, our predictions lie on top of the data points within errorbars, even in the inelastic region. This is not surprising, as inelastic channels open up slowly above kinematic thresholds.

\onecolumngrid

\twocolumngrid

\bibliography{ref}

\end{document}